\documentclass[prl,superscriptaddress,twocolumn,showpacs,
amsmath,amssymb]{revtex4}
\usepackage{units}
\usepackage{color}                   
\def \revisedt [#1]  { \colorbox{blue}{#1 }}      
\def \revisedb [#1]  { \textcolor{blue}{#1 }}   
\usepackage{graphicx}
\begin{document}

\title{Indirect Exchange Interaction in Fully Metal-Semiconductor Separated SWCNTs Revealed by ESR}

\author{M. Havlicek}
\author{W. Jantsch}
\affiliation{Johannes Kepler Universit\"at Linz, Institut f\"ur Halbleiterphysik,
 1090 Linz, Austria}
 \author{Z. Wilamowski}
\affiliation{Polish Academy of Sciences, Al. Lotnikov 32/46, PL 02-668 Warsaw, Poland}
\author{K. Yanagi}
\affiliation{Tokyo Metropolitan University,
 Tokyo, Japan}
\author{H. Kataura}
\affiliation{AIST, Tsukuba, Japan}
\author{M. H. R\"ummeli}
\affiliation{IFW Dresden,
Dresden, Germany}
\author{H. Malissa}
\author{A. Tyryshkin}
\author{S. Lyon}
\affiliation{Department of Electrical Engineering, Princeton University,
Princeton, USA}
\author{A. Chernov}
\author{H. Kuzmany}
\affiliation{Universit\"at Wien, Fakult\"at f\"ur Physik,
Strudlhofgasse 4, 1090 Wien, Austria}

\date{submitted to PRL, \today}

\begin{abstract}
  The ESR response from highly 
metal-semiconductor(M-SC) separated SWCNTs for temperatures $T$ between 0.39 and 200\;K is 
characteristically different for the two systems.  The signal originates 
from defect spins but interaction with free electrons leads to a larger line width for M tubes. The latter
decreases with increasing $T$ whereas it increases with $T$ for SC tubes. 
The spins undergo a ferromagnetic phase transition below around 10\;K. 
Indirect exchange is suggested to be responsible for the spin-spin interaction,
supported by RKKY interaction in the case of M tubes. 
For SC tubes spin-lattice relaxation via an Orbach process 
is suggested to
 determine the line width.
\end{abstract}

\pacs{73.22.-f;72.25.Rb}

\maketitle
\section{Introduction}
Electron spin resonance from single-walled carbon nanotubes (SWCNTs) has been a subject of extensive 
studies in the recent past, starting in 1997 \cite{Petit97PRB} and 
continuing until most recently a set of new papers \cite{Havlicek10pss,Dinse10AMR,Zaka10ACSN} appeared.
In several of the early papers response from localized and itinerant spins was claimed as well as 
from various collective 
phenomena such as ferromagnetic or antiferromagnetic ordering or even superconductivity
\cite{Likodimos07PRB,Corzilius07PRB,Galambos09pss}.
The discussion about the observation of spins from itinerant electrons 
entered a new state with the publication of a recent theoretical work \cite{Dora08PRL} where  
electrons in a Tomonaga-Luttinger liquid state 
(TLL) were postulated to have an extremely broad ESR absorption line, even at temperatures below 4 K. The TLL 
state was well demonstrated for SWCNT from photo emission. It is a consequence of the quasi-onedimensional 
(1D) nature of the electrons in the tubes \cite{Ishii03N,Rauf04PRL}. 
\par
The discovery of possibilities to quantitatively separate semiconducting (SC) and 
 metallic (M) tubes \cite{Yanagi08APE,Arnold06NNT} can be considered as a crucial step 
 forward to tackle the problem,
 particularly if tubes were grown from non-magnetic catalysts \cite{Ruemmeli07JPC}. 
This finding allowed for the possibility to study the ESR response separately in 
the two types of tubes. 
Preliminary results were published recently for SC tubes immersed into 
ethanol as a tube carrying medium \cite{Havlicek10pss}.  However, in this case the solvent was interacting with 
the tubes obscuring intrinsic properties. 
\par
Here we report on ESR from fully M-SC separated SWCNTs grown from a non-magnetic PtRhRe catalyst. 
Even though no response from itinerant electrons could be detected down to 0.39 K significant 
 differences in the ESR response were observed for the first time from the localized spins in the 
 two tube systems. The differences were particularly obvious 
for the line widths which were strongly enhanced at low temperatures for the M tubes  
and decreased with increasing $T$. In contrast, for the SC tubes, they were rather small at 
low temperatures but increased with $T$. 
For the ESR susceptibility
 a Curie-Weiss behavior was observed with ferromagnetic coupling. 
At $T_{\rm c}$ a ferromagnetic state is obtained, without contributions of magnetic components. 
Interaction of the spins is suggested to be by indirect exchange with an additional contribution of long range 
RKKY type interaction for the M tubes. In the latter this interaction also dominates the line width 
whereas for the SC tubes the increase of the line width with $T$ is ascribed to spin lattice 
relaxation via an Orbach process.
\par
\section{Experimental}
Single-walled carbon nanotubes with an average diameter of 1.6\;nm were grown by laser ablation 
using a PtRhRe catalyst and M-SC separated to better than 97\% as described previously 
\cite{Ruemmeli07JPC}. 
Bulk properties of the samples were characterized by transmission electron microscopy (TEM), X-ray diffraction 
and transport measurements. TEM revealed a clear bundling of the tubes with a bundle size between 20 and 
100 nm and tapered ends \cite{Supplement}. From X-ray diffraction a bundle peak at $q = 3.75$\;nm$^{-1}$ was 
observed with slightly smaller and larger values for the M and for the SC tubes, respectively. 
This corresponds to an average tube diameter of 1.6 nm in agreement with the observation from TEM. 
Conductivity measurements revealed an almost temperature independent resistivity for the metallic tubes, 
following weak localization behavior. This indicates good contact and evidences some 3D character of 
the metallic tubes. In contrast, SC tubes exhibited an exponential increase of the resistivity with 
decreasing temperature following a variable range hoping behavior \cite{Supplement}.
ESR experiments were performed at X-band frequencies of 9.45\;GHz (39.07\;$\mu$eV) 
in the temperature range between 1.5 and 200\;K with low power (30 dB) to avoid saturation. 
For temperatures between 0.39 and 1.7\;K a cylinder resonator was used built into a He3 refrigerator
and a superconducting magnet. In this case microwave power was limited to 45 dB. 
\par
Samples from 500\;$^\circ$C vacuum annealed tubes
were prepared as bucky paper with high (Pt1) and low (Pt2) thickness compared to the skin depth of 
the microwaves. The latter was evaluated from the dc conductance.
The tubes were inserted into a standard ESR quartz tube and pumped to high vacuum before measurement. 
\par 
Spin concentration was estimated by a comparison to a "strong pitch" calibration sample from Bruker. 
This sample consists of pitch embedded into a KCl matrix with a standardized 
spin concentration of 1.2$\times 10^{15}$ spins per cm length in a standard 4\;mm ESR quartz 
tube with 3\;mm inner diameter. 
Calibration was performed at 1.5\;K where the peak to peak line width is 1.7\;G. 
Considering the reduction of cavity quality factor by the nanotube samples 
we obtained 7.5$\times 10^{17}$\;spins/g for the M tubes and 12$\times 10^{17}$\;spins/g for the SC tubes.
Within experimental error this is the same spin concentration and results in 1 spin per 1000 carbon atoms. 
\par
\begin{figure}
\includegraphics[width=0.8\linewidth]{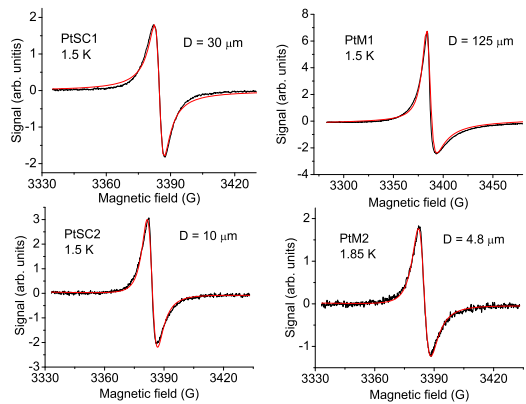}
\caption{Derivative of ESR absorption for bucky paper samples of different thickness $D$ as indicated.
Left side graphs: SC tubes, right side graphs: M tubes.
The smooth lines in the lower graphs are fits
using Eq.\;\ref{eq:Dyson} with $g= 2.0043$. For the fits of the upper graphs the anomalous 
skin effect had to be considered \cite{Feher55PR,Supplement}.}
\label{fig:ESRline}
\end{figure}

The ESR spectra were fitted considering a dispersive signal contribution. 
Samples with high thickness behave textbook like as the M tubes show a well expressed asymmetry in the 
absorption whereas the absorption in the SC tubes remains nearly symmetric. Results for low temperature 
recording are depicted in Fig.\;\ref{fig:ESRline}. 
For the case of the low thickness samples the ESR response is nearly symmetric. 
The remaining slight asymmetry 
was found to be well represented by the derivative of a Dysonian line shape of the form
\begin{equation}
 \frac{\partial \chi}{\partial B}= \frac{-2A(B-B_0) w\cos \phi}{(w^2+(B-B_0)^2)^2}+\frac{(w^2-(B-B_0)^2)A\sin \phi}{(w^2+(B-B_0)^2)^2}\;,
\label{eq:Dyson}
\end{equation}
where $A, B_0, w, $ and $\phi$ are fitting parameters describing the signal amplitude, 
the resonance field, the line width, and a phase angle. For $\phi \ne 0$ a contribution of the dispersive component 
of the susceptibility to the absorption line appears.
Since the analysis for the low thickness samples is much simpler, in the following results are reported only for 
the latter. The ESR susceptibility was evaluated as $\chi \propto Aw^2$. 
 \par
\section{Results}
According to Ref.\;\onlinecite{Dora08PRL} the line widths for the itinerant electrons should be around 1000\;G at 4\;K
for a TLL parameter $K_{\rm s} = 1.1$. For a very small decrease of $K_{\rm s}$ due to bundling
of only metallic tubes
and for temperatures as low as 0.4\;K 
the line width should reduce to about 50\;G, well within the range observable in our experiment. 
However, even at the lowest temperature of 0.39 K no extra broad line is observed with a  
free carrier response signature. 
\par
\begin{figure}
\includegraphics[width=\linewidth]{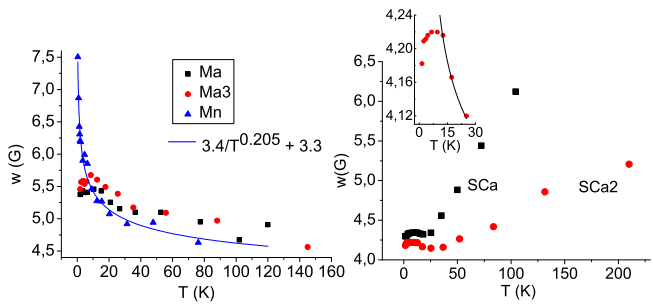}
\caption{ESR line width for M  (left) and SC (right) tubes. Blue triangles: unannealed tubes, 
black squares and red bullets: two different 
annealed tubes.  Blue line: inverse power law fit to the triangles as indicated. SC tubes: 
two examples for annealed tubes. Insert: blow up for SCa2 at low temperature. The full drawn line is an 
inverse power law fit.}
\label{fig:MSClw}
\end{figure}

For the narrow lines depicted in Fig. \ref{fig:ESRline} we 
observe characteristic differences in the line widths between SC and M tubes with respect to their 
dependence on temperature and with respect to their values at low temperatures.
The line width increases with decreasing temperature for the M tubes but decreases for the SC tubes. 
As a consequence, at low temperatures the line width for the former is significantly larger 
as compared to the line width of the latter. 
Examples are depicted in Fig.~\ref{fig:MSClw}.
The increase of the line width for low temperatures is particularly dramatic for the M tubes before annealing.
The fit with 
an inverse power law yields 0.205 for the exponent and a constant {\it residual line width} of 3.3\;G. 
The observed increase in line width ({\it excess line width}) is more than a factor two from its  
residual width.
For the annealed tubes a similar power law 
behavior is obtained for high and intermediate temperatures but the increase stops and levels off 
around 10\;K. In all cases the negative exponent is 0.2$\pm 0.05$. 
\par
For SC tubes the line width increases with temperature at least for high and intermediate temperatures, 
first in a quasi-linear manner with a trend to saturate for higher temperatures. 
The total increase amounts up to 2\;G for the temperature range under study.   The initial slope varies with 
sample treatment and decreases with tube annealing.
At temperatures between 20 and 30\;K the line width becomes almost temperature independent with a noticeable
 re-increase for $T\rightarrow 0$ and a peak around 8\;K. 
 Due to the strong signal at these temperatures 
 the line widths were determined with a RMS error of $\pm 0.009$ G which is approximately two times the size of the 
 red dots in the insert. 
\par
\begin{figure}
\includegraphics[width=\linewidth]{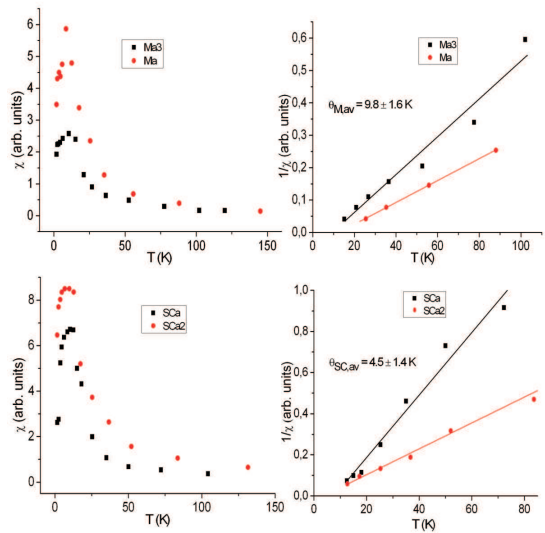}
\caption{Temperature dependence of ESR susceptibility $\chi \propto Aw^2$ 
for two samples as evaluated from amplitude $A$ 
and line width $w$ for M (upper graphs) 
and SC tubes (lower graphs), respectively. Right part: $1/\chi$ 
vs $T$ in each case, together with a linear regression. 
$\theta_{\rm M,av}$ and $\theta_{\rm SC,av}$: averaged intersections of the 
regressions with the temperature axis.} 
\label{fig:temp1}
\end{figure}

In addition to the line width, the magnetic susceptibility was evaluated from the ESR signal as a 
function of temperature.
The temperature dependence of $\chi$ exhibits a maximum value around 10\;K but 
is $1/T$ 
like for higher temperatures. As a consequence at high temperatures the lines are hardly detectable but 
recorded with high precision at low temperatures. 
 Figure \ref{fig:temp1} depicts examples for 
M and SC tubes.  
The peak in the ESR susceptibility coincides well with the leveling of the line width for the M samples 
at low temperatures and with the re-increase of the line width for the SC samples.
Using a 
Curie-Weiss law for $\chi$ as $\chi \propto 1/(T-\theta)$ renders the spin-spin 
interaction as ferromagnetic rather than anti-ferromagnetic as the linear regressions in Fig.\;\ref{fig:temp1}
intersect the temperature axis at positive values. The observed peak in $\chi(T)$ suggests a transition to a 
ferromagnetic state with a finite distribution of transition temperatures.
\par
\section{Discussion}
 The lack of the observation of additional ESR contributions from free carriers 
 is consistent with
  a full TLL state of the electrons even in the bundled metallic tubes with relaxed 
 1D character. It is also consistent with recent experiments 
from photo emission where for metallic tubes TLL signatures were observed in the spectral 
function \cite{Ayala10PRB}. 
\par
The localized spins can be assigned to various defects such as single bonded covalent 
functionalization or defects on or inside the tubes. The defect concentration is similar to the 
value of $2.5\times10^{-3}$/C reported in \cite{Corzilius07PRB} from magnetic measurements but 
considerably higher than values of $1.5\times 10^{-6}$ given in \cite{Likodimos07PRB}. 
The high spin concentration observed here is not surprising since both, purification and 
separation introduce defects. 
This is also evidenced from Raman scattering which revealed an increase of the D/G ratio by 
33\% from 0.06 to 0.08 (515 nm laser) for the SC tubes and by 57.5\% from 0.13 to 0.21 (647 nm laser) 
for the M tubes. 
The defects are in strong interaction with the tubes leading in general to similar behavior for  
M and SC tubes. However, as far as transition temperatures and spin relaxation 
are concerned, significant differences are observed. 
\par
Wave functions of the defect spins are in general spread out over a considerable number 
of hexagonal cells. For example, DFT calculations for hydrogen 
bonded to a small capped (3,3) 
SWCNT yield a spin distribution over 5 and 8 hexagons in axial and radial direction, respectively
\cite{Bulusheva10unp}.
This provides the possibility of exchange interaction in spite of the large distance between spins. 
The interaction can be particularly enhanced by 
the indirect exchange mechanism across the carbon lattice,
and, for M tubes, by interaction with free carriers according to the RKKY mechanism. 
\par
In the suggested ferromagnetic state we do not observe the ESR response from the magnetically ordered spins,
since the easy axes of the magnetic regions are randomly oriented following the random orientation of the 
nanotubes or bundles of nanotubes. 
Magnetic anisotropy can be traced back to an anisotropic curvature perpendicular and parallel to the tube axis 
and the concomitant anisotropy in the SO coupling. 
\par 
Magnetic properties from defects in carbon systems have been reported in several publications 
\cite{Makarova04SeCon,Wang09NL,Yang09ACSN}. 
Apparently transversal extension of the tubes or the 3D nature of the bundles
is enough to allow ferromagnetic or anti-ferromagnetic interaction but it is not enough to suppress TLL nature 
of the electrons.
A decrease of spin susceptibility at low temperatures was recently reported for unseparated SWCNTs 
\cite{Likodimos07PRB} and speculated to originate from a spin gap opening or from an anti-ferromagnetic ordering 
of the defect spins as it was suggested for the magnetic behavior of nanohorns \cite{Garaj00PRB}. 
The low concentration of defect spins in \cite{Likodimos07PRB} may be a problem for such interpretation. 
In contrast, in our case with the much higher spin concentration the temperature dependence of the 
inverse susceptibility exhibits clear ferromagnetic coupling for SC and M tubes. 
Indications for weak ferromagnetism in small parts of nanotube samples were also reported from 
magnetization measurements of SWCNT with high concentration of defect spins in \cite{Corzilius07PRB}. 
However in this case it was 
suggested that ferromagnetic ordering occurs in metallic tubes only.
\par
In the M tubes the 
transition temperatures $\theta_{\rm M}$ as evaluated from the linear regressions 
are almost a factor two higher than in the SC tubes. This is suggested to be due to the additional 
spin-spin interaction mediated by the free carriers.
In low dimensional electronic systems the
RKKY type interaction is very efficient since it extends as $1/R^D$ where $R$ is the distance between two spins
 and $D$ is the dimension of the system \cite{Shenoy05PRB}. In SWCNT it is 
 particularly interesting since Fermi wave vectors
   can be either zero or $2\pi/3G$ where $G$ is the NT translation vector \cite{Costa05PRB}.
   As a consequence the RKKY interaction
   is not oscillatory, at least not for tubes with $k_{\rm F}=0$. RKKY interaction is 
likewise important for the transition temperatures and for the line widths 
and was shown in \cite{Costa05PRB} to decrease with increasing $T$.
\par
\begin{figure}
\includegraphics[width=\linewidth]{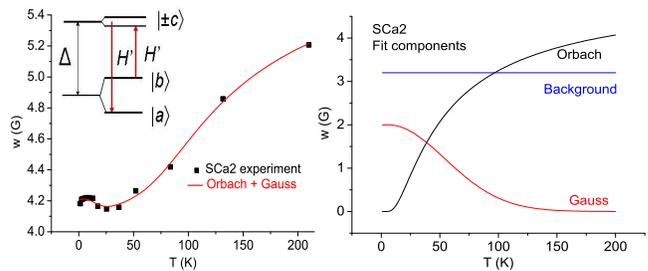}
\caption{ESR line width $w$ for semiconducting samples versus temperature $T$. 
Squares: Line width from the experiment, line: fit 
according to Eq.\;\ref{eq:OrbGauss} (left) and explicit representation of the 
fit components (right).  Fit parameters were $A = 5.1$\;G, $\Delta = 45.1$\;K for the Orbach 
relaxation and $A_{\rm g} = 2.7$\;G, $T_{\rm g} = 5.2$\;K, $w_{\rm g} = 69$\;K for the 
Gaussian distribution, and $w_0=3.2$\;G for the background. 
The insert shows the scheme for the Orbach process.}
\label{fig:wfit}
\end{figure}

The increase of the line width with increasing temperature for SC tubes suggests spin-lattice 
interaction (SLI) as the relevant 
process for spin relaxation.  SLI is usually mediated by either a direct process which is linear in $T$, 
a Raman process which goes with a high power of $T$ or an Orbach process. Since only the latter exhibits 
saturation with $T$ it is natural to invoke it for the description of the experiments. 
On the other hand, the re-increase of the line width at very low temperatures indicates additional contributions. 
Since it is suggestive that this increase is related to the ferromagnetic phase transition
with the statistically distributed transition temperatures 
a Gaussian temperature dependence is appropriate for its description. 
Considering homogeneous and inhomogeneous contributions we describe the line width obtained from 
the Dysonian fits as
\begin{equation}
w^2 = \Delta_{\rm L}^2(T) + \sigma_{\rm G}^2(T)+ w_0^2\,,
\label{eq:OrbGauss}
\end{equation}
where $\Delta_{\rm L}$ is a Lorentzian type homogeneous line width following an Orbach relaxation 
$\Delta_{\rm L}=A \exp (-\Delta/T)$
and $\sigma_{\rm G}=A_{\rm g}\exp(-(T-T_{\rm g})^2/w_{\rm g}^2)$
considers the line width from Gaussian fluctuations and possible concomitant variations of $g$ 
values. $w_0$ is a temperature independent background line width. This Pythagorayan 
addition of line width components is a commonly used approximation which mimics a folding procedure 
\cite{Simon06PRL}.
$\Delta$ and $A$ in the expression for the Orbach process are the energetic distance to 
 a higher electronic state for the spin excitation, and a scaling prefactor. 
$A_{\rm g}$, $T_{\rm g}$, and $w_{\rm g}$ are the parameters of the distribution from inhomogeneous 
fluctuations. As shown in Fig.\;\ref{fig:wfit},  Eq.\;\ref{eq:OrbGauss} provides a very good 
representation of the experiments 
with $A$, $\Delta$, $A_{\rm g}$, $T_{\rm g}$\, $w_{\rm g}$, and $w_0$ as parameters.
From the parameter values in the caption of Fig.\;\ref{fig:wfit} the excitation energy is 3.8 meV (45.1\;K). 
Applying the same fitting procedure to the other sample gave very similar values. 
$\Delta$ was in particular found to be 44.5\;K for the second sample. 
\par
In the case of a spin-lattice relaxation by an Orbach process the excited spin state $|b\rangle$ 
and the ground state 
$|a\rangle$ are coupled by a perturbation Hamiltonian $H'$ to an intermediate split off electronic 
state $|\pm c\rangle$ 
to which spins are excited before relaxation. A scheme of this process is depicted in 
Fig.\;\ref{fig:wfit}. The perturbation is in general phononic or 
vibronic. In the original work of R. Orbach \cite{Orbach61PRS} the extra electronic 
state came from crystal field splitting. 
Due to the large number of defects in our samples a rather large number of
possible excited states can be expected which supports a spin-lattice relaxation by an Orbach process. 
\par
The increase of the line widths for $T \le 20$\;K in the SC tubes 
is the signature of fluctuations when approaching the phase transition. 
The values from Fig.\;\ref{fig:temp1} for $T_{\rm c} = 4.5 \pm 1.4$\;K are
in good agreement with the value of  $T_{\rm g}=5.2$\;K used for the fit in Fig.\;\ref{fig:wfit}. 
Fluctuations along with structural phase transitions are well known 
to broaden ESR lines. \cite{Blinc96AMR,Mozurkewich84PRB}
This broadening often 
follows a power law of the form $a/(T-T_{\rm c})^m$ with $m$ of the order of 1.
The fit depicted in the insert of Fig.\;\ref{fig:MSClw} yields $m=0.22$, slightly smaller than expected.
This deviation is very likely due to the statistical distribution of $T_{\rm c}$ 
with the concomitant flattening of the slope.

\par
For the metallic tubes a decrease of line width with increasing temperature is observed.
 Since in our case 
the defect spins are localized and do not show activated behavior motional narrowing can be ruled out 
for an explanation. 
Increasing and even diverging line widths for decreasing $T$ were also reported for metallic spin systems 
which undergo a transition to a spin glass \cite{Mozurkewich84PRB}. 
 However, in this case line widths of the order of 100\;G are expected.
Since this is not observed here we are left with an indirect exchange mechanism as responsible for the 
increase of line with decreasing temperature. In the metallic tubes this mechanism 
has a strong contribution from the free carrier RKKY interaction which increases with decreasing temperature
as described above.
\par
One may certainly ask why spin lattice interaction e.g. in the form of an Orbach process is not 
active for the metallic tubes. Such behavior can indeed not be ruled out. As it would partly 
compensate the negative temperature coefficient observed for the metallic tubes such relaxation 
would result in an even stronger decrease of line width with temperature for the metallic tubes.
\par
The surprise in the results for the line width concerns not only the different 
temperature dependence but even more the considerably larger line width for the M tubes as compared to the SC 
tubes at low temperatures. This, together with the different temperature dependence, leads to a 
crossover of the line widths with increasing temperature. 
While the ratio between the excess line width for the M tubes and for the SC tubes is 3.2  at 1.5\;K on average, it is 
only 0.6 at 200\;K. This is suggested to be a consequence of the increasing spin-lattice interaction with increasing 
$T$ and the simultaneous decrease of the line width due to the decreasing RKKY interaction.
\par
\section{Summary}
In summary we have shown that M-SC separated SWCNT reveal significant differences in their ESR response. 
While for both types of tubes a ferromagnetic temperature dependence is observed for the spin 
susceptibility,  the latter levels off at characteristic temperatures $\theta_{\rm {SC, M}}$ 
and eventually decreases. This is interpreted as a transition to a ferromagnetic state. 
The difference in the transition temperature for the two tube systems 
is suggested to originate from an RKKY type interaction of the spins with the free carriers in the M tubes. 
Even more dramatic is the observed difference in the temperature dependence of the line width for 
the two tube systems. It decreases with increasing temperature for the M tubes but increases for 
the SC tubes. 
Since the behavior of the metallic tubes follows the 
temperature dependence of the RKKY interaction the latter is suggested to be responsible for the line 
narrowing. In contrast, for SC tubes an Orbach type relaxation process is suggested to be 
responsible for the broadening of the 
ESR lines width increasing temperature. 
\par
\begin{acknowledgments}
  Valuable discussions with B. Dora, F. Simon, M. Mehring, and K. Baberschke are
  gratefully acknowledged. Work supported by FWF projects P17345 and P20550, A and
   DFG project RU1540/8-1, D.
\end{acknowledgments}

\end{document}